\documentstyle[prb,aps,twocolumn]{revtex}

\begin{document}

\draft
\flushbottom
\twocolumn[\hsize\textwidth\columnwidth\hsize\csname @twocolumnfalse\endcsname

\title{Analysis of atomic-orbital basis sets from the projection of
plane-wave results}

\author{Daniel S\'anchez-Portal, Emilio Artacho, and Jos\'e M. Soler}
\address{
   Instituto de Ciencia de Materiales Nicol\'as Cabrera and
   Departamento de F\'{\i}sica de la Materia Condensada, C-III
\\
   Universidad Aut\'onoma de Madrid, 28049 Madrid, Spain.}
\date{\today}
\maketitle

\begin{abstract}
The projection of the eigenfunctions obtained in standard
plane-wave first-principle calculations is used for analyzing
atomic-orbital basis sets. The "spilling" defining the error in
such a projection allows the evaluation of the quality
of an atomic-orbital basis set for a given system and its systematic
variational optimization. The same projection allows obtaining the
band structure and the Hamiltonian matrix elements in the
previously optimized atomic basis. The spilling is shown to correlate
with the mean square error in the energy bands, indicating that the
basis optimization via spilling minimization is qualitatively equivalent
to the energy-minimization scheme, but involves a much smaller
computational effort. The spilling minimization also allows the
optimization of bases for uses different than total-energy calculations,
like the description of the band gap in semiconductors.
The method is applied to the characterization of finite-range
pseudo-atomic orbitals {[}O. F. Sankey and D. J. Niklewski, Phys. Rev. B
{\bf 40}, 3979 (1989){]} in comparison to infinite-range pseudo-atomic
and Slater-type orbitals. The bases are evaluated and optimized for
several zincblende semiconductors and for aluminum. The quality of the
finite-range orbitals is found to be perfectly comparable to the
others with the advantage of a limited range of interacions.
A simple scheme is proposed to expand the basis without increasing
the range. It is found that a double-$z$ basis substantially
improves the basis performance on diamond, whereas $d$-polarization
is required for Si and Al for similar results. Finally, the projection
allows the chemical analysis of the plane-wave results via population
analysis on the previously optimized atomic basis. The spilling can
partially control the intrinsic arbitraryness of the population analysis.
\end{abstract}
%
\pacs{ }
]

\section {INTRODUCTION}
One of the several approximations needed for electronic structure calculations
of solids is the truncation of the one-particle Hilbert space.
Choosing an appropiate basis is critical for obtaining high-quality results.
There are schemes based on localized, extended, or mixed bases.
Most of the methods of the first group use atomic-like bases,
the eigenfunctions being obtained as linear combinations of atomic
orbitals (LCAO methods).\cite{eschrig} The most widely used among the extended
is the plane-wave basis.\cite{cohen}

Plane waves (PW) together with pseudopotentials have shown
to offer a very successful calculation scheme for a very large number of
applications. Particulary since the development of {\it ab initio}
pseudopotentials\cite {hamann} PW have provided the tool for
very accurate calculation of different properties of solids.\cite{cohen}
PW are quite simple to implement and the convergence
of the calculations can be controlled with a single parameter,
the planewave cut-off. The PW basis is also independent of atomic
positions, which is very convenient for the coding. However, in addition
to these advantages, plane waves have important drawbacks, namely
(i) the imposed translational invariance (supercells for non-periodic systems),
and (ii) their inefficiency in basis size.

LCAO methods are much more efficient in the size of the required basis.
This is a very important advantage for the calculation of large systems.
Moreover, they have shown to be very suitable for order-N
methods\cite{ordern} in which the computational effort scales
linearly with the size of the system. However, this large reduction
in the size of the basis is obviously accompanied by a potential loss of
completeness which can affect the results. It is then necessary
to choose an appropiate basis to obtain accurate results.
Nevertheless, it is important to stress that, once a high-quality atomic
basis set has been chosen, calculations can be performed with an accuracy
perfectly comparable to that of plane waves and with a large reduction
of CPU time and memory.

Several methods have been developed to optimize LCAO basis sets.
Most of them are based in two procedures: ({\it i}) minimization
of total energies of atoms, molecules, or solids,\cite {poirier}
and ({\it ii}) minimization of differences in the energy bands
comparing with the experiment or with PW calculations.\cite{eschrig,chadi}
We present here a projection method that links PW and LCAO schemes
and provides: ({\it i}) a systematic way to characterize the quality of an
atomic basis and therefore a way to optimize it, ({\it ii}) LCAO energy
bands and Hamiltonian and overlap matrix elements,
and ({\it iii}) chemical information through population analysis.
The key of the optimization method is to measure the ability of a set of
atomic orbitals to represent the actual eigenstates of the PW
calculation of a system. By minimizing the error in the projection
of the eigenstates the atomic basis can be optimized for that
particular system. This scheme is related to methods of quantum chemistry
where a relatively small LCAO basis set is optimized by minimizing
the distance (maximizing the overlap) of its subspace to the one
spanned by another, larger LCAO basis.\cite{MOM,pelos} The scheme used
in this paper was already outlined in Ref.~\onlinecite{SSC}, the projection
technique for bands being already sketched by Chadi in Ref.~\onlinecite{chadi}.

Besides the projection optimization, the possibility of calculating LCAO
band structures and matrix elements allows the characterization of
(i) the energetics, and (ii) the range of interactions of the basis.
Particulary for this purpose non-orthogonalized atomic basis sets
are more adequate, since the interaction parameters have proved to
be much more transferable and to have shorter range than for
orthogonal bases.\cite{porfy,eafy} The direct use of atomic
orbitals in condensed systems gives non-orthogonal basis sets naturally.
These non-orthogonal atomic basis sets will be used throughout this
paper.\cite{emilio}

In the present paper, this method of analysis and optimization is
applied to the characterization of basis sets made of the finte-range
pseudo-atomic orbitals introduced by Sankey and Niklewski.\cite{sankey}
They are compared with the infinite-range pseudo-atomic
orbitals\cite{sankey2} and Slater-type orbitals. Both minimal
and expanded bases are explored as applied to several zincblende
semiconductors, as well as to aluminum.

The optimized basis can be used for the calculation of larger systems,
where plane-wave calculations are impractical. Besides that possible use,
they can also be used directly for the analysis of the PW results
themselves.\cite{SSC} The chemical language of population analysis is
not accesible
to the PW scheme. The projection into the optimum LCAO basis allows
its application, to compute atomic charges, bond charges, and charge
transfers. In a similar way as the decomposition of the total charge
into orbital populations, the density of states can also be projected
into atomic orbitals to give local densities of states.
It is shown here how the optimization of the basis is important
for obtaining sensible values of the charges, and also how the
projection can be used to partially control the intrinsic arbitraryness
of any population analysis.

The structure of this paper is as follows. A brief description of
the method to characterize and optimize the atomic basis
is given in section II, as well as a description of the way
of performing the population analisys and the band structure
calculations. Section III describes the main characteristics
of the {\it ab initio} PW calculations used here. A description of
the different types of atomic orbitals analyzed in this paper can be
found in section IV. Section V proposes the spilling as a parameter
to characterize basis sets and their correlations with other quantities.
Section VI contains the results of the analisys and optimization
of the differents sets of atomic orbitals. Population analysis is
discussed in section VII. We end with the conclusions
in section VIII.

\section { DESCRIPTION OF THE METHOD}

The method presented in this paper is based on projection
techniques.\cite{chadi,SSC}
Given the results of a PW calculation, the projection of the
calculated eigenstates into an atomic basis can provide information
about this basis set and about the PW results. A good basis must be able
to represent
the essential features of the PW eigenstates. Once the PW calculation
has been performed for the system of interest, the projection process
is much less costly than the self-consistent calculation itself, and many
different basis sets can be analyzed and optimized for one single PW
calculation, a much more economic procedure than performing a full
self-consistent calculation for each trial basis set.

On the base of our method we have the definition of the {\it spilling}
parameter.\cite{SSC} Given a PW calculation we define the spilling
${\cal S}_{\Omega}$ for a given local basis set as
\begin{equation}
\label{deviation}
{\cal S}_{\Omega}={1\over N_k} {1\over N_\alpha} \sum_{k}^{N_k}
\sum_{\alpha=1}^{N_\alpha} \langle \psi_\alpha ({\bf k}) | (1-P({\bf k}))
| \psi_\alpha ({\bf k}) \rangle ,
\end {equation}
where $|\psi_\alpha ({\bf k}) \rangle $ are the PW calculated eigenstates,
and $N_k$ and $N_\alpha$ are the number of calculated {\bf k} points in the
Brillouin zone and the number of bands considered, respectively, and
$_{\Omega}$ is the specific subspace spanned by the eigenstates
included in the sum (defined by $N_k$ and $N_\alpha$). $P({\bf k})$
is the projector operator into the subspace of Bloch functions of wave vector
{\bf k}, generated by the atomic basis, and defined as usual for non-orthogonal
basis,\cite {emilio}
\begin {equation}
\label{projector}
P({\bf k})= \sum_{\mu} |\phi_\mu({\bf k}) \rangle \langle \phi^\mu({\bf k})| =
\sum_{\mu \nu} |\phi_\mu({\bf k}) \rangle S^{-1}_{ \mu \nu} ({\bf k})
\langle \phi_\nu({\bf k}) | ,
\end {equation}
where
\begin{equation}
\langle {\bf r} | \phi_\mu({\bf k}) \rangle =
\sum_{\bf R} \phi_\mu({\bf r}-{\bf r}_\mu-{\bf R})
e^{i{\bf k}({\bf r}_\mu+{\bf R})}
\end {equation}
$\phi_\mu({\bf r})$ being the atomic orbitals, ${\bf r}_\mu$ the atomic
coordinates in the unit cell, and $ {\bf R}$ the lattice vectors.
\begin{equation}
S_{ \mu \nu}({\bf k})=\langle \phi_\mu({\bf k})|\phi_\nu({\bf k}) \rangle
\end{equation}
is the overlap matrix of the atomic basis. The vectors
$|\phi^\mu({\bf k}) \rangle$ constitute the dual of the atomic basis, that
satisfy
\begin{equation}
\langle \phi^\mu({\bf k})|\phi_\nu({\bf k}) \rangle =
\langle \phi_\mu({\bf k})|\phi^\nu({\bf k}) \rangle =
\delta_{\mu \nu}.
\end{equation}
${\cal S}_{\Omega}$ is a parameter that varies between 0 and 1 and
measures the  ability of a basis to represent PW eigenstates, by
measuring how much
of the subspace of the Hamiltonian eigenstates falls outside the
subspace spanned by the atomic basis.
If we consider the PW eigenfunctions
$|\psi_\alpha ({\bf k}) \rangle $ and their
projection into the atomic basis
$P({\bf k})|\psi_\alpha ({\bf k}) \rangle $, then
${\cal S}_{\Omega}$ gives the average of $|| (1-P)|\psi_\alpha ({\bf k})
\rangle ||^2$ over the eigenstates considered for the projection.
It must be also pointed out that the projected eigenstates
$| \chi_\alpha ({\bf k}) \rangle = P({\bf k})|\psi_\alpha ({\bf k})\rangle$
are not necessarily orthonormal, the overlap matrix
$R_{\alpha \beta}({\bf k})=\langle \chi_\alpha ({\bf k})| \chi_\beta ({\bf k})
\rangle$ being different from the identity matrix. In particular
$R_{\alpha \alpha}({\bf k}) \leq 1$, and
\begin{equation}
{\cal S}_{\Omega}={1\over N_k} {1\over N_\alpha} \sum_{k}^{N_k}
\sum_{\alpha=1}^{N_\alpha} (1-R_{\alpha \alpha}({\bf k}))
\end {equation}
is the averaged norm which has been lost (spilled) in the
projection process. If we restrict our analisys to the {\it occupied}
eigenstates, then ${\cal S}_{\Omega}$ is the fraction of total electronic
charge spilled when projecting, and will be refered to as {\it charge
spilling} ${\cal S}_Q$.

A local spilling can also be defined as
a function of real space coordinates,
\begin {equation}
\label {spfunction}
{\cal S}_{\Omega}({\bf r})={1\over N_k}
{1\over N_\alpha} \sum_{k}^{N_k} \sum_{\alpha=1}^{N_\alpha}
(|\psi_\alpha ({\bf k},
{\bf r})|^2 - |\chi_\alpha ({\bf k},{\bf r})|^2 )
\end {equation}
We consider it a useful and direct way for visually characterizing
the quality of the representation of the PW eigenstates.
The total spilling defined above is recovered by integrating the
local spilling over the three-dimensional space.

Important for the characterization of the quality of a basis for a given
system is the energy band structure. To calculate the LCAO energy bands of
a system we project the Hamiltonian associated to the self-consistent
PW charge density of our reference calculation into the basis subspace:
\begin {equation}
\label {hamiltonian}
H^{LCAO}_{\mu \nu}({\bf k})= \langle \phi_{\mu}({\bf k})|H^{PW}|
\phi_{\nu}({\bf k}) \rangle
\end {equation}
To obtain the LCAO Hamiltonian matrix elements
we expand the Bloch basis functions into plane waves.\cite{chadi}
We calculate the kinetic energy and the effect of the non-local
part of the pseudopotential directly in Fourier space.
The effect of the local part of the potential and self-consistent
Hartree and LDA
interactions is calculated using fast Fourier transform algorithms,
\begin {equation}
\label {hamiltonian2}
H^{LCAO}_{\mu \nu}({\bf k})=\!\!\!\!\!\!\!\sum_{|{\bf k+G}|^2<E_{max}}
\!\!\!\!\!\!\!\!\!\! \langle
\phi_{\mu}({\bf k})|{\bf k + G}\rangle\langle{\bf k + G}|H^{PW}|
\phi_{\nu}({\bf k}) \rangle
\end {equation}
\begin {equation}
\label {Blochoverlap}
S^{LCAO}_{\mu \nu}({\bf k})=\!\!\!\!\!\!\!\sum_{|{\bf k+G}|^2<E_{max}}
\langle \phi_{\mu}({\bf k})|{\bf k + G} \rangle \langle
{\bf k + G}| \phi_{\nu}({\bf k}) \rangle
\end {equation}
where $\bf G$ are reciprocal lattice vectors and $E_{max}$ is an energy
cut-off independent of the one used in the PW calculation, that must be
large enough to guarantee a reliable representation of the atomic basis.
With this method we obtain a LCAO Hamiltonian from a first-principles PW
calculation, no free parameters being fitted.

The LCAO Hamiltonian is obtained in Bloch space and, as a consequence,
it includes interactions up to infinite neighbors.
The deviations in the band structure with respect to the PW
only reflect the uncompleteness and inadequacy of the basis,
clearly separated from the possible additional approximation of
neglecting matrix elements beyond some range of interactions. We
can study both approximations separately. In most
studies about LCAO band structures these two effects are mixed together.
As an exception it must be mentioned the works of Chadi\cite{chadi} and
Jansen and Sankey,\cite{sankey2} where also LCAO Hamiltonians with
interactions up to an infinite range of neighbors are obtained.

The Hamiltonian matrix in real space is calculated from the Hamiltonian
matrix in Bloch space by performing a Bloch type transformation
\begin {equation}
\label {parameter}
H^{LCAO}_{\mu \nu}({\bf R}_{\mu \nu})=\sum_{\bf k} H^{LCAO}_{\mu \nu}
({\bf k}) e^{i{\bf k}({\bf R}_\mu - {\bf R }_\nu)}
\end {equation}
\begin {equation}
\label {parameteroverlap}
S^{LCAO}_{\mu \nu}({\bf R}_{\mu \nu})=\sum_{\bf k} S^{LCAO}_{\mu \nu}
({\bf k}) e^{i{\bf k}({\bf R}_\mu - {\bf R }_\nu)}
\end {equation}
where normalization factors which depend on the overlaps are omitted for
clarity. The sum has to be extended to a sufficient number of
{\bf k} points, taking into account that the number of points depends
on the real space range of the interactions.\cite{monkhorst,josepoints}

The projection technique can also be used for obtaining chemical information
from the PW calculations by means of an LCAO
population analisys. We use the one proposed by Mulliken.\cite{mulliken}
The PW occupied eigenstates are projected onto the subspace spanned by
the atomic basis. Due to the non-orthogonality of the projected eigenstates,
the density operator has to be properly defined to ensure the charge
conservation
\begin {equation}
\label{density operator}
\hat \rho = \sum_{{\bf k}} \sum_{\alpha}^{occ} |\chi_\alpha({\bf k})\rangle
\langle \chi^\alpha ({\bf k}) | \, ,
\end {equation}
where $|\chi^\alpha ({\bf k})\rangle= \sum_{\beta} R_{\beta \alpha}^{-1}
({\bf k})| \chi_\beta ({\bf k})\rangle$ represent the vectors of the dual
set of the projected eigenstates, and
$R_{\beta \alpha}({\bf k})=\langle \chi_\beta ({\bf k}) |
\chi_\alpha ({\bf k})\rangle$. The LCAO density matrix is then written
in terms of the dual LCAO basis:
\begin {equation}
\label{density matrix}
{\cal P}_{\mu \nu}=\langle \phi^{\mu}|\hat \rho|\phi^{\nu} \rangle \, ,
\end {equation}
the charge associated to an orbital $\mu$ being
\begin {equation}
\label{charge of an orbital}
Q_\mu=\sum_\nu {\cal P}_{\mu \nu } S_{\nu \mu} .
\end {equation}
\section {REFERENCE CALCULATIONS}
The analysis of the bases presented in this paper are based on reference PW
calculations which have been performed using
{\it ab initio} pseudopotentials generated by the improved
Troullier-Martins method\cite {TM} and within the local density
approximation (LDA) for electron exchange and
correlation.\cite{lda} Of course, other schemes could have been
equally used, the only requirement to apply the method being
the planewave expansion of the one-electron wavefunctions.
The energy cut-off
for the plane-wave expansion has been taken large enough to
consider that all the calculations are well converged.
High energy cut-offs are not necessary for the application
of the method but are important to ensure accurate results for
the optimization of the orbitals as well as for the population analisys.
The cutoffs used are: 10 Ry for Al, 20 Ry for
Si, BP, and AlP, 30 Ry for GaAs, 40 Ry for graphite, 50 Ry for BN,
and 70 Ry for C and SiC. For the self-consistent zincblende
semiconductor calculations, we used 2 irreducible-Brillouin-zone
points (IBZP), equivalent to 32 whole-Brillouin-zone points (BZP), giving
a k-mesh {\it length cutoff} \cite{josepoints} $l_c=a$, where $a$ is the
lattice constant. For aluminum and graphite we used respectively (10 IBZP,
125 BZP, $l_c$=13.435 a.u.) and (8 IBZP, 16 BZP, $l_c$=4.648 a.u.).
The length cutoff $l_c$ is simply related with the maximum range of the
real-space matrix elements obtainable from the PW calculation
(Eqs. 11 and 12) by $|{\bf R}_{\mu \nu}| < l_c $.
For the pseudopotentials, in all the calculations we have used the
separable, fully non-local formulation following Kleinman and
Bylander.\cite{KB}

\section {ATOMIC BASIS SETS}

A brief description follows of the sets of atomic orbitals
which are used throughout this work.
The studied basis sets can be classified in two groups:
({\it i}) those which are provided by the numerical solution of
the atomic problem (with the same atomic pseudopotential
as used in the PW calculation), hereafter called PAO's
for pseudo-atomic orbitals,\cite{sankey,sankey2}
and ({\it ii}) Slater-type orbitals (STO).\cite{poirier}

Within the first group, PAO$_{\infty}$ are the atomic solutions under
real atomic boundary conditions (at $r=\infty$).
This family of bases allows for a scaling factor for the radial part
acting as a variational parameter to optimize
the atomic orbitals via the spilling minimization,

\begin{equation}
\phi_l^{opt} ({\bf r})= \lambda_l^{3\over 2} \phi_l^{atm}(\lambda_l{\bf r}) .
\end{equation}

PAO$_{\infty}$ represents quite a natural choice, since access to PW
calculations is usually accompanied by access to atomic calculations
under the same
approximations. They provide a good description of the charge density
and have already been used in {\it ab initio} calculations\cite{sankey2}
giving accurate results. For covalent materials, with small charge
transfers, they provide a very good description of the pseudo-core region.
This is worth considering since the PW eigenstates are expanded in the
atomic basis maximizing the projection without making any distinction
between the pseudo-core region (zone near the atomic nucleus)
and the valence region (interatomic zone).

The main disadvantage of the PAO$_{\infty}$'s is the very long range
of the interactions and overlaps they originate. Getting a shorter
 and perfectly defined range of interactions
 is the reason to work with PAO$_{r_c}$.
They are the solutions of the atomic pseudopotential when
forced to vanish outside a cutoff radius $r_c$, keeping
the continuity, as introduced by Sankey and Niklewski\cite{sankey}
(they will be denoted by specifying the value of $r_c$ in atomic units).
The cutoff radius represents a variational parameter for this type of bases.
These orbitals
seem promising candidates in calculation techniques
for very large systems.\cite{ordern}

Within the Slater orbitals, we call STO1 the
conventional orbitals $r^{n-1} \exp^{-\beta r}$, where $n$
is the principal quantum number and the scale factor $\beta$
is taken to be the same for all the atomic-orbitals within the same shell;
STO2 will be the same, but allowing for a different $\beta$ for
each orbital, $\beta_s\neq\beta_p\neq\beta_d$; and
STO3 are the orbitals $r^{\alpha} \exp^{-\beta r}$, where
$\alpha$ becomes another variational parameter, no longer restricted to
be an integer, and allowing both $\alpha$ and $\beta$ to be different
for each orbital.

STO's have been extensively used in LCAO calculations
in quantum chemistry and have proven to provide accurate results.\cite{poirier}
They present the additional advantage of their very simple
mathematical properties.  To simplify the computational task,
it is a common practice to
expand the STO's orbitals as linear combinations of gaussian functions.
These are the so-called STO-NG\cite{poirier} orbitals where N stands for the
number of gaussians orbitals used per STO. We have done some tests to
compare the results obtained using STO and STO-4G orbitals.
No appreciable differences have been found in the results. Spilling
values and optimum exponents are essentially the same in both cases,
and no further investigation has been pursued concerning STO-NG.

STO orbitals do not seem {\it a priori} to be very well suited
for the description of the pseudo-core region for the projection,
possibly leading to worsen the spilling at the valence region while
trying to adapt the pseudo-core. To explore that possibility
we consider a last kind of orbital functions, the pseudo-STO,
which represent a link between the STO and the PAO families.
A pseudo-STO is a Slater-type orbital outside a certain
radius, and the numerical solution of the pseudopotential inside that
radius, continuity of the function and its derivative being imposed.

\section {SPILLING ANALYSIS}

This section is devoted to show the adequacy of an analysis
based on the projection and the spilling for the evaluation of a
basis set. The total spilling provides a natural parameter that
can be used straightforwardly to measure systematically the quality
of a basis. The optimization in terms of spilling provides a
variational scheme, since the minimum attainable is zero, which
means a perfect basis.

Both the projected charge density and the local spilling
offer a useful tool for visually characterizing the quality of a basis.
This is illustrated in Fig. 1, where a charge-density contour map for SiC is
shown comparing the PW results with the results of the projection
on the PAO$_{\infty}$, showing in addition the local spilling
along the bond axis. We see that  some of the bond structure is lost,
the charge density being much more spherical around the atoms, when
projecting on this basis.
This can be also seen in the fact that the two maxima in the density
near the carbon nucleus are of approximately the same height,
while in the PW calculation there is more density at the bond side.
All these features can be seen in the local spilling, which
is negative in the regions with an excess of projected charge
and positive in the zones with a defect of projected charge.
The local charge spilling is negative in the proximity of both
core regions and positive in the bond region. Some of the
bond charge goes to an anti-bonding region. This feature
also tends to make the charge-transfer lower for this PAO$_{\infty}$
basis (see below).

For the spilling analysis to be useful, the spilling
has to be correlated with the quality in the energetics of the system.
This is tested in Fig. 2, where the mean square error of the projected
valence bands of crystralline silicon, as compared to the PW ones, is plotted
versus the charge-spilling ${\cal S}_Q$, for different families of bases.
It is calculated without shifting the bands, the absolute zero being
unambiguously defined (the Hamiltonian operator is not changed).
This quantity can also be taken as an indicator of
the error in the total energy due to the approximation of the basis.

The clear correlation between the spilling and the error in the
energy bands displayed in Fig. 2 indicates that a basis optimization
based on a minimization of the spilling is parallel (at least
approximately) to an energy minimization. This parallelism is
even clearer within a basis family, where the correlation
is much more pronounced (different points within the same family are
obtained by varying the characteristic parameters presented above).

To finish this section, we introduce
a lower bound for the spilling.
 The dependence of ${\cal S}_{\Omega}$ on the atomic basis enters through the
Fourier coefficients of the radial wave-functions. Therefore,
${\cal S}_{\Omega}$ can be regarded as a function of these Fourier
coefficients $\phi_{i,\mu}(|{\bf k}+{\bf G}|)$ which are different for each
atom $i$, orbital $\mu$ and each modulus $|{\bf k}+{\bf G}|$. Considering
all these Fourier coefficients as free varables the minimization of
${\cal S}_{\Omega}$ gives a lower
bound for ${\cal S}_{\Omega}$. Values of this lower bound will be shown
as Min. ${\cal S}$ in the tables.

\section {ANALYSIS AND OPTIMIZATION OF ATOMIC BASIS SETS}

The quality of the LCAO bases presented above is measured and optimized
for different solids in this section using the spilling analysis, and,
simultaneously, the LCAO band structure is contrasted with the
results of plane-waves. The results are displayed in Tables I through IV.
The resulting numbers are abundant since the number of possibilities is
large. For the different kinds of bases introduced in section IV there
are different possible basis sizes, single-$z$ or double-$z$, meaning one
or two orbitals per atomic symmetry. There is the additional possibility
of polarizing the basis, i.e., introducing atomic orbitals of a higher angular
momentum than needed, which, in the cases of this paper represents the
addition of the $d$ orbitals to the minimal $sp$ basis.

The optimization of the different bases introduces new degrees of freedom
since their optimization is done with respect to different variational
parameters (see section IV), and also the spilling can be defined for
different numbers of bands, depending on the particular application the basis
is optimized for.

\subsection{Minimal basis sets}

Table I shows the spillings of the projections onto different minimal
bases for silicon, diamond, and aluminum, together with the mean
square error in the LCAO bands as compared with the PW ones. Only $sp$ bases
are considered, except for Al, for which also $spd$ bases data
are shown due to the poorness of minimal $sp$ bases for this solid.
The numbers displayed for STO bases have been obtained after minimization
of the charge spilling (the spilling of the valence eigenvectors) for
the corresponding parameters. This procedure is illustrated in Fig.~3 for
the specific cases for silicon and carbon, where the charge spilling is
shown as a function of the STO exponents. PAO$_{\infty}^{opt}$ stands
in Table I for the PAO$_{\infty}$ optimized via a scale factor.
The PAO$_{r_c}$ are optimized with respect to the radius of the
basis functions,$r_c$, as illustrated in Figs.~4, 5, and 6 for
silicon, diamond, and aluminum, respectively. In Table I, especific
values of $r_c$ are chosen.

All the bases displayed in Table I perform comparably well in what
concerns the charge spilling. Clearly better does $sp$ for C than
for Si and Al, as expected since $d$ orbitals play a lesser role for C
than for Si and Al. The spilling for eigenstates including conduction
bands is larger than for valence bands only. For $sp$ bases there is
a factor of 10 to 20. The two main reasons for that are, (i) the bases
have been optimized by minimizing the charge spilling, and (ii) higher
energy bands require higher energy atomic orbitals.

Table I also shows the populations of $s$, $p$, and $d$ orbitals
(if present). It can be seen that the $sp^3$ hybridization is
better accomplished in C than in Si (a perfect hybridization would
show $Q_s=1$  and $Q_p=3$). $Q_d$ in Al is quite appreciable.

The mean square error in the energy bands is displayed as
$\sigma_v$ for the valence bands and $\sigma_c$ for the first two
conduction bands ($\sigma_{1+2}$ for first two bands, and $\sigma_{3+4}$
for third and forth, in the case of aluminum). The valence bands are
quite well described with $sp$ bases for Si and C, but not for Al.
The metallicity in Al demands a larger contribution of $d$ orbitals,
what makes them necessary for any quantitative LCAO description of Al.
The conduction bands are much worse described by the minimal bases
shown in the table.

The dependence in the number of bands to be optimized through the
spilling is shown in Table II for Si, C, BN, and Al. Each kind of basis
has been optimized by minimizing (i) the charge spilling and (ii)
a spilling including conduction bands (${\cal S}_8$ for Si, C, and BN,
and ${\cal S}_4$ for Al). It can be seen how an improvement in
${\cal S}_{8(4)}$ is accomplished when optimizing the basis for it,
but paying the prize in the worsening of ${\cal S}_Q$.
In addition to the ${\cal S}$ and
$\sigma$ values, the band gap region description is evaluated by showing
the values of the $\Gamma$-direct and indirect gaps, and the position
of the minimum of the conduction band. An important and systematic
improvement is observed in these quantities when the optimization is
extended to unoccupied states. The displayed information offers the
necessary information for the evaluation of the trade-off in the
quality of the different spectral regions for the choice of a basis
for a particular application.  Fig.~7 shows the band structure of Si
obtained with a minimal $sp$ STO3 basis optimized considering
${\cal S}_8$. Notice the nice agreement in the
band-gap region, showing a very realistic indirect gap, quite rare
for a minimal basis [compare with the typical cases shown in Fig.~8 (a)
and (d)]. The LCAO bands have been shifted by 1.04 eV, what means
an important deviation in total energies, if computed. Another price
to pay is the very long range interactions due to the unusual extension
acquired by the orbitals in the optimization.

Trying to improve the performance, other, more complicated, kinds of
bases have also been tried, the results not being shown in the Tables.
Pseudo-STO's (see section IV) provide good bases, but the improvement
on the others (STO and PAO) have not been substantial enough for
seriously considering them as alternative candidates. They
have been optimized by varying the STO parameters outside the core region
(as for STO3) plus the extra degree of freedom given by the core radius.
For Si a minimum in ${\cal S}_Q$ is obtained for a core radius of 1 a.u.,
but the spilling is only 4 \% lower than the spilling of the STO3 case.
This fact indicates that the projection is not as sensitive to the
core regions so as to worsen the overall projection if the core region
in the basis orbitals is not adapted to the pseudopotential.

Another kind of basis has been constructed based on PAO$_{r_c}$. More
degrees of freedom are given to the spilling minimization by linear
combining excited states of the atomic problem to the ground state.
For each symmetry ($s,p_x,p_y,p_z,...$) the atomic orbital is constructed
with a linear combination of the ground and excited states of the
corresponding pseudopotential. This procedure considerably reduces the
spilling without increasing the size of the basis. However, the band
structure features
are not accordingly improved. This fact leads to the conclusion that,
for the spilling-energy correlation to be satisfactory in the solid,
the energetics in the basis itself (in the atomic problem) has to be
considered.

\subsection{Range of interactions}

An important aspect defining the applicablity of LCAO bases is their
range of interactions, i.e., the scope of neighbor shells that have to
be included in the LCAO Hamiltonian and overlap matrices. The basis sets
giving comparable results in the previously shown quantities perform quite
differently in this respect. This is shown in Table III.
This information is obtained by computing the atomic matrix elements
of the Hamiltonian and overlap matrices as described in section II.
Due to the different analytical expressions for the orbitals, to
measure the interaction range we adopt an empirical criterium, namely,
the radius $r_{99.9\%}$ for which 99.9 \% of the norm of the orbital
is within the sphere with that radius. We have
found that criterium meaningful in the sense that neglecting interactions
beyond that radius does not appreciably alter the energy bands corresponding
to the valence and some conduction bands (average error lower than 0.05 eV).
Neglecting interactions within
that radius produces appreciable deviations in the bands.
For PAO$_{r_c}$'s the interacting scope is clearly defined by $r_c$ (which
incudes 100 \% if the norm within), but for short $r_c$'s they are
very similar to $r_{99.9\%}$, and for larger $r_c$'s then $r_{99.9\%}$ becomes
more meaningful. Therefore, we also characterize the PAO$_{r_c}$'s with
$r_{99.9\%}$. Besides this quantity, Table III also shows the scope of
non-negligible neighboring shells for each case.

PAO$_\infty$ represents the best basis in what spilling and band quality is
concerned, and also for population purposes\cite{SSC} (see next section),
but its range of interactions is very large. This fact was
already pointed out by Jansen and Sankey\cite{sankey2} in their work
doing self-consistent calculations using this basis. The results of
our calculations show that interactions up to at least 5 (10) shells
of neighbors have to be kept for Si (C) to obtain a reasonable band
structure with the correct semiconducting behavior.
The best basis suited for a shorter range of interactions together
with keeping good standards in spilling and band structure is the
PAO$_{r_c}$, as can be seen in Table III and Fig.~4 for Si, in
Fig.~5 for C, and in Fig.~6 for Al. Note that, in spite of C needing a
smaller $r_c$ than Si to attain small spillings, the range of neighbors
is larger for C due to the much smaller lattice parameter.

\subsection{Polarized and double-$z$ basis sets}

The clearest way of improving the quality of a basis is increasing its
size. In LCAO there are two usual ways of doing it, (i) increasing the
number of orbitals within the same atomic symmetries as found in the
minimal basis, and (ii) polarizing the basis. Within the first, double-$z$
is the most usual case, i.e., doubling the basis, which in the cases
of this paper means a basis of two $s$ ,two $p_x$, two $p_y$ and two $p_z$
orbitals.
A polarization of the basis is accomplished when a shell of different
atomic symmetry is added to the minimal basis. In our cases it means
the addition of the $d$ shell to the minimal $sp$ basis. Combination of
these two procedures is also customary for further improvement. A rule
of thumb in quantum chemistry says that a basis should always be doubled
before being polarized.\cite{poirier} This is not the usual practice
within the LCAO calculations in the solid-state-physics community, where
double-$z$ bases are hardly to be found.

Expanded STO bases can be found in the literature.\cite{poirier} This
is not the case for PAO's. In this paper we propose to use the excited
states of the atomic calculations leading to the PAO$_{r_c}$'s, under
the same boundary conditions, to obtain the orbitals required to expand
the basis, i.e., excited $s$ and $p$ orbitals for double-$z$ bases and
$d$ orbitals for polarization. The advantages of this scheme are
its simplicity and the controlled range of interactions also for
the added orbitals, which is usually lost in the traditional scheme.

Again, we study here the quality of expanded LCAO bases by means of
their spilling and their band structure as obtained from the projection.
The results of our analysis are shown in Table IV, where minimal (single-$z$),
double-$z$, and polarized single-$z$ bases are compared for silicon and
diamond. See also Table~I for aluminum. The $sp$ and $spd$ bases are
optimized as above using STO3 for the Slater orbitals, and using the
corresponding $r_c$ for the PAO's, as shown in Figs.~4-6. Double-$z$
for STO has been taken from the literature.\cite{poirier,STO2-z} Note
that, with double-$z$, Si does not reach equivalent quality as C, and
that, for that purpose, the polarization of the basis is quite enough.

The band structures corresponding to the data of Table IV are shown
in Fig.~8 for Si, Fig.~9 for C, and Fig.~10 for Al (Table~I), in each
case comparing the LCAO bands obtained from the projection with the
reference PW bands. The information given in the Table is ratified, and
observed in more detail. Note that the substantial improvement in the Si
case occurs when including the $d$ orbitals, but not when doubling the
minimal basis. For diamond, however, such qualitative improvement is
given by the doble-$z$ basis. The figures for the gap in the Si and
C cases are shown in Table V. In the Al case, the polarization is needed
from the start, the $sp$ projection giving defective band structures in
the neighborhood of the Fermi level around the W point. The $spd$ band
structure is shown in Fig.~10 for PAO$_{r_c}$ for different $r_c$.
The performance of the double and polarized STO bases in the first
conduction bands is better than that of the corresponding PAO$_{r_c}$
bases. This is due to the larger extension of the extra orbitals in
the STO bases, which is not allowed in the PAO$_{r-c}$'s.
The latter, however, also represent a considerable improvement over the
minimal bases, while keeping the interaction range. This virtue makes them
very appealing for accurate LCAO calculations.

The error in the band structures could be noticeably reduced with a rigid
shift of the bands, which is clear in sight of the Figures. This
has not been allowed in order to keep the band-structure error as
representative of the behavior to be expected of total energies.

\subsection{Geometry and environment dependence}

The projection procedure proposed in this paper is based on PW calculations
of actual materials. As compared to basis optimization in atoms, this
allows adequating the basis to particular environments or geometries.
As a first example we show the change in the minimal $sp$ PAO$_{r_c}$
basis for C depending on the solid being diamond or graphite. Fig.~11
shows the charge spilling as a function of $r_c$ and of the range of
interacting neighboring shells. Graphite is worse represented
by the same kind of basis. In addition, graphite requires larger $r_c$ to
approach minimum spilling. This latter fact is due to the comparatively
large interplanar distances in the graphite structure.

Another example of geometry dependence of the basis quality is given in
Fig.~12, where the charge spilling of a minimal PAO$_{r_c}$ for Si is plotted
versus the lattice constant for different values of $r_c$. From the
Figure it is clear that (i) for spilling purposes and, therefore,for
the total energy, the basis
is better the larger $r_c$, (ii) the slope in ${\cal S}_Q$ versus lattice
constant will translate in a corresponding slope of the total energy,
what originates a fictitous pressure on the system, (iii) this pressure
is different depending on $r_c$, and the optimum $r_c$ for this purpose
(the one giving zero pressure, with $r_c$ between 5 and 6 a. u.) is different
from the optimum for energy considerations ($r_c=\infty$), and (iv) there
are no appreciable
curvatures, from where it can be inferred that no substantial deviations
are to be expected in the ellastic and force constants (vibrations) due
to the basis.

\section{POPULATION ANALYSIS}

One of the uses of the projection into LCAO is the possibility of
performing population analysis of the PW results, otherwise inaccesible.
It is well known that any population
analisys has an inherent arbitraryness associated to the fact that the
relevant quantities in the analysis do not correspond to observables,
but rather to convenient ways of dissecting the total charge of the
system. Furthermore, small variations in the basis can produce
quite different population results, particulary for charge transfers.
In spite of it, much of the chemical descriptions of condensed systems
is performed in the language given by such population analyses. We,
therefore, consider it worthwhile to explore the problem with the spilling.

Due to the mentioned arbitraryness, the correlation of charge transfers
and charge spilling is expected to be poor. This is clear when
uncompensating the basis, like, for instance, when introducing
$d$-orbitals for silicon in SiC. This greatly improves
the description of the eigenfunctions and reduces the spilling,
but being the basis uncompensated, some features (structure
of the bond) are described by the much more complete basis of
one of the atoms, so that more electronic charge is assigned to the
atom with more complete basis, getting an unphysical charge transfer.
A compensated basis is therefore important for the populations to be
meaningful. A measure of this compensation can be obtained if the
spilling is calculated independently for structurally similar solids
containing the individual atoms separately. The spillings for silicon
and diamond separately, using the same kind of basis for both, indicate a
slight
uncompensation favoring diamond, which, in this case says that the
charge transfer might be slightly overestimated. In addition to that,
the optimization of the basis is important to get reliable values
of charge transfers. This is seen in Fig.~13, where the charge transfer
in SiC is shown to correlate with the spilling. A 'converged' charge
transfer is not guaranteed, but, provided that the basis is compensated,
a small spilling indicates a meaningful charge transfer.

Table VI shows the charge transfers for some zincblende semiconductors,
obtained with the projection procedure, using optimized PAO$_{\infty}$
basis sets. The numbers compare well with equivalent results obtained
from direct LCAO selfconsistent calculations.\cite{orlando,GaAs}
Systems with only one species do not present these difficulties for the
charges since the basis can be trivially compensated. Table VII shows
the excess of charge (with respect to the neutral atom) assigned to the
upper and lower atoms of a dimer in the Si(100)-2$\times$1 surface.\cite{Si2x1}
A minimal basis has been used. The populations are quite stable and
independent of the basis used.

\section{CONCLUSIONS}

We have presented a projection method which provides a tool to
characterize and optimize atomic orbital basis sets for a given
system. Independently of how many bases are evaluated for that
particular system,  only {\it one} reference self-consistent planewave
calculation is required. The method also provides LCAO band structures,
Hamiltonian matrix elements (non fitted tight-binding parameters), and
population analysis.

Based on that tool and on its characteristic parameter, the spilling,
an exhaustive analysis of two kinds of LCAO basis sets, namely, Slater-type
(STO) and pseudoatomic (PAO) orbitals,  have been presented for several
solids, arriving to the following conclusions:

($i$) The spilling provides a very convenient tool for
studying and variationally optimizing basis sets, that closely
correlates with energy optimizations.

($ii$) The LCAO band structures obtained from the projection allow
to characterize separately the errors due to the uncompleteness of the basis
and to considering a limited range of interactions. This is due to the fact
that the bands are obtained directly in Bloch space, with interactions
to infinite neighbors.

($iii$) Both STO's and PAO's optimized with spilling minimization
give quite good basis sets, of comparable quality.

($iv$) If a controlled range of neighbor interactions is
desired, the best performing basis sets are provided by the PAO$_{r_c}$
introduced by Sankey and Niklewski.\cite{sankey}

($v$) A simple and interaction-range-controlled scheme for extending
PAO's basis sets to double-$z$ and/or polarized bases is proposed, and
shown to give good basis sets.

($vi$) Double-$z$ bases substantially improve the performance on diamond,
whereas $d$ polarization is required for similar results in Si and Al.

($vii$) The spilling minimization allows optimization taylored for
particular uses, like for the description of excitations or band gaps,
environment dependence of the basis, etc.

($viii$) Population analysis can be performed, and its intrinsic
arbitraryness can be partially controlled by means of the spilling.

\acknowledgments

We acknowledge useful discussions with F. Yndur\'ain and P. Ordej\'on.
This work has been supported by the Direcci\'on General de Investigaci\'on
Cient\'{\i}fica  y Tecnol\'ogica of Spain (DGICYT) under grant PB92-0169.

%


\begin {figure}
\caption{Electron charge density for SiC: (a) self-consistent density
contour map in the (110) plane calculated with PW, (b) the same after
projecting the eigenstates into a PAO$_{\infty}$ $sp$ basis, and (c)
these two densities (PW with solid lines and the projected
density with dotted lines) and
${\cal S}_Q ({\bf r})$ along the bond axis. Units are electrons per unit cell.
}
\end{figure}

\begin{figure}
\caption{Mean square deviation in valence bands of the projected hamiltonian
from the PW bands versus the charge spilling of different bases in
silicon. Each symbol stands for a different family of bases: $\Box$
for PAO$_{r_c}$, $\times$ for PAO$_\infty$, $+$ for STO, and $\Diamond$
for different double-$z$ bases. Bases with $\sigma_v$ lower than
0.2 eV include $d$ orbitals.}
\end{figure}

\begin{figure}
\caption{Charge spilling versus exponent for a STO1 basis set,
(a) silicon with a {\it spd} basis, (b) diamond with a {\it sp}
basis. Also shown the optimum exponents for a STO2 basis set for
the same systems.}
\end {figure}

\begin {figure}
\caption{(a) Charge spilling for silicon with a PAO$_{r_c}$ {\it sp}
and {\it spd} basis versus $r_c$, (b) charge spilling versus
the shell of neighbors with non-zero overlaps for the {\it spd}
basis, and (c) for the {\it sp} basis.}
\end{figure}

\begin{figure}
\caption{(a) Spilling for occupied and first eight bands for
diamond  with a PAO$_{r_c}$ {\it sp} versus $r_c$, (b) charge-spilling
versus the shell of neighbors with non-zero overlaps
basis, (c) the same for the spilling of the first eight bands.}
\end{figure}


\begin{figure}
\caption{The same as Fig.~4 for aluminum.}
\end{figure}

\begin {figure}
\caption{Gap optimization with a minimal basis. Silicon band
structure calculated with a STO3 {\it sp} optimized to reproduce
the occupied eigenstates and the first four conduction bands.
The PW band structure is shown with dots. The STO3 bands
have been shifted 1.04 eV to make the top of both valence bands coincide.}
\end {figure}

\begin{figure}
\caption{Band structure of Si for minimal, double-$z$, and $spd$ bases.
In (a), (b) and (c) the Hamiltonian is projected in a STO3 $sp$, STO
double-$z$, and STO3 $spd$ basis, respectively. STO3 bases optimized
for valence band. (d),(e) and (f) correspond
to PAO$_{5.0}$  $sp$, double-$z$, and $spd$ basis. The PW reference band
structure is shown with dots.}
\end {figure}

\begin {figure}
\caption{Band structure of diamond for single-$z$ and double-$z$ bases.
(a) and (b) correspond the projection in a STO3 $sp$
and STO double-$z$ basis sets, respectively. STO3 basis optimized for
valence band. Figures (c) and (d)
correspond to PAO$_{4.0}$ $sp$ and double-$z$.
The PW reference band structure is shown with dots.}
\end {figure}

\begin {figure}
\caption{Aluminum band structure: (a) calculated with PAO$_{5.3}$ $spd$
basis, and (b) using PAO$_{6.1}$. Dotted lines show the PW band structure
(Fermi energy is 8.2 eV).}
\end {figure}

\begin {figure}
\caption{(a) Charge spilling for diamond (D) and graphite (G) as a function
of $r_c$ for a PAO$_{r_c}$ $sp$ basis. (b) Charge spilling
for diamond as  a function of the shell of neighbors with
non-zero overlap, and (c) the same for graphite.}
\end{figure}

\begin{figure}
\caption{(a) ${\cal S}_Q$ versus lattice constant of silicon for
PAO$_{r_c}$ bases, (b) $\delta${\cal S}$_Q$={\cal S}$_Q$-{\cal S}$_Q^{eq}$
where the {\cal S}$_Q^{eq}$ is taken at the equilibrium value of the
lattice constant $a_0$=10.2 a.u.. $\Diamond$ stands for
r$_{c}$=4 a.u., $+$ for r$_c$=5 a.u., $\Box$ for r$_{c}$=6 a.u., $\times$
for r$_{c}=\infty$.}
\end{figure}

\begin {figure}
\caption{Charge transfer in SiC as a function of the charge spilling
of the {\it sp} basis used to project the density.
$\Box$, stands for PAO$_\infty$ optimized for the heteropolar system
and independently for Si and C in a zinc-blende structure. $\Diamond$,
stands for STO3 basis in the same conditions, and $+$ for a STO2 basis.}
\end{figure}

\newpage



\begin{table}
\caption{Bases for silicon ($sp$), diamond($sp$), and aluminum
($sp$ and $spd$). ${\cal S}_{Q}$
stands for the charge spilling, and ${\cal S}_{4}$ and ${\cal S}_{8}$
for the spilling considering the first 4 and 8 bands, respectively.
PAO$_{\infty}$ and PAO$_{r_c}$ are fixed bases, whereas STO1, STO2, STO3,
and PAO$_{\infty}^{opt}$ have been optimized according to their respective
variational parameters, to minimize ${\cal S}_Q$.
$Q_s$, $Q_p$, and $Q_d$ are the total charges assigned to $s$, $p$,
and $d$ orbitals. The mean square error of the LCAO bands with
respect to the PW reference bands (in eV) is displayed for the valence bands
($\sigma_{v}$) and two first conduction bands ($\sigma_{c}$) for silicon
and diamond, and for the first and second bands ($\sigma_{1+2}$) and
third and forth ($\sigma_{3+4}$) for aluminum. Min ${\cal S}$ corresponds to
the spilling lower bound (see text).}
\vspace{2pt}
\begin{tabular}{llllllllll}
& Basis & & ${\cal S}_{Q}$ & ${\cal S}_{8}$ & $Q_s$ & $Q_p$ & &
$\sigma_{v}$ &  $\sigma_{c}$ \\
\hline
\\
   & PAO$_{\infty}$ &$sp$& 0.0080 & 0.1596 & 1.32 & 2.68 & & 0.30 & 2.15 \\
   & PAO$_{\infty}^{opt}$&$sp$& 0.0078 & 0.1621 & 1.35 & 2.65 & & 0.29 & 1.97\\
   & PAO$_{5.0}$&$sp$& 0.0139 & 0.1591 & 1.40 & 2.60 & & 0.36 & 2.09 \\ \\
Si & STO1 &$sp$& 0.0109 & 0.1764 & 1.48 & 2.52 & & 0.38 & 2.43 \\
   & STO2 &$sp$& 0.0076 & 0.1562 & 1.36 & 2.64 & & 0.34 & 2.34 \\
   & STO3 &$sp$& 0.0074 & 0.1548 & 1.34 & 2.66 & & 0.29 & 1.95 \\ \\
   & Min ${\cal S}$&$sp$ & 0.0044 & 0.0689 \\
   & Min ${\cal S}$&$spd$ & 0.0001 & 0.0014 \\
\hline \\
& Basis & & ${\cal S}_{Q}$ & ${\cal S}_{8}$ & $Q_s$ & $Q_p$ & &
$\sigma_{v}$ &  $\sigma_{c}$ \\
\hline
\\
   & PAO$_{\infty}$ &$sp$& 0.0035 & 0.0719 & 1.05 & 2.95 & & 0.32 & 2.19 \\
   & PAO$_{\infty}^{opt}$ &$sp$&0.0027&0.0780&1.07 & 2.93 & & 0.25 & 2.61 \\
   & PAO$_{4.0}$ &$sp$& 0.0041 & 0.0672 & 1.10 & 2.90 & & 0.23 & 2.22 \\ \\
C  & STO1 &$sp$& 0.0039 & 0.0771 & 1.11 & 2.89 & & 0.26 & 3.07 \\
   & STO2 &$sp$& 0.0039 & 0.0775 & 1.12 & 2.88 & & 0.24 & 3.05 \\
   & STO3 &$sp$& 0.0024 & 0.0790 & 1.09 & 2.98 & & 0.17 & 3.36 \\ \\
   & Min ${\cal S}$&$sp$ & 0.0008 & 0.0226 \\
\hline \\
& Basis & & ${\cal S}_{Q}$ & ${\cal S}_{4}$ & $Q_s$ & $Q_p$ & $Q_d$ &
$\sigma_{1+2}$ &  $\sigma_{3+4}$ \\
\hline
\\
  &PAO$_{\infty}$&$sp$&0.0165 & 0.1141 & 1.13 & 2.87 &  & 2.03 & 3.45\\
  &PAO$_{\infty}$&$spd$&0.0022 & 0.0053 & 0.77 & 1.39 & 0.84 & 0.61 & 0.24\\
  &PAO$_{\infty}^{opt}$&$spd$&0.0011&0.0079& 0.98&1.41&0.61 & 0.35 & 0.84\\ \\
  &PAO$_{6.0}$ &$sp$& 0.0157 & 0.1012 & 1.30 & 1.70 &   & 1.54 & 2.97\\
Al&PAO$_{6.0}$ &$spd$& 0.0014 & 0.0042 & 1.02 & 1.39 & 0.59 & 0.20 & 0.45\\
  &PAO$_{5.3}$ &$sp$& 0.0158 & 0.0930 & 1.34 & 1.66 &  & 1.34 & 2.63\\
  &PAO$_{5.3}$ &$spd$& 0.0007 & 0.0020 & 1.15 & 1.45 & 0.40 & 0.97 & 2.15\\ \\
  &Min ${\cal S}$ &$sp$& 0.0009 & 0.0025 \\
  &Min ${\cal S}$ &$spd$& 2.9 10$^{-5}$ & 0.0007 \\
\end {tabular}
\end{table}


\begin{table}
\caption{Basis optimizations for different number of bands, for Si,
C (diamond), BN, and Al. Superindex indicates the number of bands used
for the ${\cal S}$ minimization ($v$ means the valence bands). The basis
is minimal, except for Al, for which it is $spd$.
${\cal S}_{Q}$ is the charge spilling, and ${\cal S}_{4}$
and ${\cal S}_{8}$ are the spilling considering the first 4 and 8
bands, respectively. The mean square error of the LCAO bands with
respect to the PW reference bands (in eV) is displayed for the valence bands
($\sigma_{v}$) and two first conduction bands ($\sigma_{c}$) for Si, C,
and BN, and for the first and second bands ($\sigma_{1+2}$) and
third and forth ($\sigma_{3+4}$) for Al.
$E_{gap}^{\Gamma}$ and $E_{gap}$ are the direct
gap at $\Gamma$ and the absolute gap, respectively,
both in eV. ${\bf k}$ is the wave vector of the minimum of the
conduction band in units of $\pi/a$, being $|{\bf k}|=1$ for the X point.}
\vspace{2pt}
\begin{tabular}{lllllllll}
& Basis & ${\cal S}_{Q}$ & ${\cal S}_{8}$ & $\sigma_{v}$ &
$\sigma_{c}$ & $E_{gap}^{\Gamma}$ & $E_{gap}$ & $|{\bf k}|$\\
\hline
\\
  &PAO$_{\infty}^{v}$ & 0.0078 & 0.1621 & 0.28 & 1.97 & 2.91 & 1.69 & 0.63 \\
  &PAO$_{\infty}^{8}$ & 0.0348 & 0.1233 & 0.94 & 2.13 & 2.33 & 1.44 & 0.80\\ \\
  &PAO$_{5.0}^{v}$ & 0.0136 & 0.1453 & 0.44 & 2.16 & 2.78 & 1.96 & 0.62 \\
  &PAO$_{5.0}^{8}$ & 0.0230 & 0.1170 & 0.91  & 2.47 &  2.50 & 0.95 & 0.83\\ \\
  &STO1$^{v}$ & 0.0109 & 0.1764 & 0.38 & 2.43 & 2.83 & 2.60 & 0.47 \\
Si&STO1$^{8}$ & 0.0207 & 0.1541 & 0.60 & 2.27 & 2.40 & 1.94 & 0.60\\ \\
  &STO2$^{v}$ & 0.0076 & 0.1562 & 0.34 & 2.34 & 2.66 & 2.09 & 0.53 \\
  &STO2$^{8}$ & 0.0306 & 0.1179 & 0.79 & 1.48 & 2.39 & 0.85 & 0.80\\ \\
  &STO3$^{v}$ &0.0074 & 0.1548 & 0.29 & 1.95 & 2.66 & 1.74 & 0.60 \\
  &STO3$^{8}$ & 0.0465 & 0.1121 & 1.20 & 1.28 & 2.52 & 0.53 & 0.83\\ \\
  &PW & & & & & 2.55 & 0.45 & 0.85 \\
\hline \\
& Basis & ${\cal S}_{Q}$ & ${\cal S}_{8}$ & $\sigma_{v}$ &
$\sigma_{c}$ & $E_{gap}^{\Gamma}$ & $E_{gap}$ & $|{\bf k}|$ \\
\hline
\\
  &PAO$_{\infty}^{v}$ & 0.0027 & 0.0780 & 0.25 & 2.61 & 6.57 & 5.83 & 0.50 \\
  &PAO$_{\infty}^{8}$ & 0.0192 & 0.0600 & 0.89 & 2.04 & 5.19 & 4.13 & 0.67\\ \\
  &STO1$^{v}$ & 0.0039 & 0.0771 & 0.26 & 3.07 & 5.78 & 5.78 & 0.00\\
  &STO1$^{8}$ & 0.0208 & 0.0592 & 0.94 & 1.95 & 4.67 & 4.34 & 0.42\\ \\
C &STO2$^{v}$ & 0.0039 & 0.0775 & 0.24 & 3.05 & 5.79 & 5.79 & 0.00 \\
  &STO2$^{8}$ & 0.0293 & 0.0545 & 1.37 & 1.88 & 4.49 & 3.51 & 0.58\\ \\
  &STO3$^{v}$ & 0.0024 & 0.0790 & 0.17 & 3.36 & 6.32 & 6.18 & 0.42 \\
  &STO3$^{8}$ & 0.0335 & 0.0526 & 1.32 & 1.46 & 4.81 & 3.96 & 0.67\\ \\
  &PW & & & & & 5.28 & 3.43 & 0.76 \\
\hline \\
& Basis & ${\cal S}_{Q}$ & ${\cal S}_{8}$ & $\sigma_{v}$ &
$\sigma_{c}$ & $E_{gap}^{\Gamma}$ & $E_{gap}$ & $|{\bf k}|$ \\
\hline \\
  &PAO$_{\infty}^{v}$ & 0.0022 & 0.0766 & 0.24 & 3.06 & 10.90 & 8.56 & 0.93 \\
BN&PAO$_{\infty}^{8}$ & 0.0154 & 0.0534 & 0.96  & 2.21 & 7.93 & 4.63 & 0.93\\
  &PW & & & & & 8.19 & 3.82 & 0.93 \\
\hline \\
& Basis & ${\cal S}_{Q}$ & ${\cal S}_{4}$ & $\sigma_{1+2}$&$\sigma_{3+4}$\\
\hline \\
Al&PAO$_{\infty}^{v}$&0.0011&0.0079&0.35 & 0.84\\
  &PAO$_{\infty}^{4}$&0.0036&0.0024&0.37 & 0.23 \\
\end {tabular}
\end{table}


\begin{table}
\caption{Interaction range for minimal $sp$ bases in Si. NS stands for
the number of neighbor shells with non-negligible interactions and
$r_{99.9\%}$ is a measure of the size of the orbitals as defined in
the text. ${\cal S}_{Q}$ stands for the charge spilling, and ${\cal S}_{8}$
for the spilling considering the first 8 bands. The mean square
error of the LCAO bands with respect to the PW reference bands (in eV)
is displayed for the valence bands ($\sigma_{v}$) and two first
conduction bands ($\sigma_{c}$). STO1$_{1.75}$ and STO1$_{1.85}$ are the
STO1 bases for the exponents
$\beta=1.75$ a.u.$^{-1}$ (Ref.~5) and 1.85, respectively.}
\vspace{2pt}
\begin{tabular}{lllllll}
Basis & ${\cal S}_{Q}$ & ${\cal S}_{8}$ & $\sigma_v$ & $\sigma_c$ &
$r_{99.9\%}$ &NS\\
\hline \\
STO1$_{1.75}$ & 0.0319 & 0.2279 & 0.96 & 4.27 & 5.1 & 4 \\
STO1$_{1.85}$ & 0.0514 & 0.2533 & 1.07 & 4.64 & 4.8 & 3 \\ \\
PAO$_{\infty}$ & 0.0080 & 0.1596 & 0.30 & 2.15 & 7.9 & 9 \\
PAO$_{5.0}$     & 0.0139 & 0.1591 & 0.36 & 2.09 & 4.6 & 3 \\
PAO$_{4.0}$     & 0.0506 & 0.2269 & 1.01 & 3.89 & 3.7 & 2 \\
\end {tabular}
\end{table}


\begin{table}
\caption{Single-$z$, double-$z$ (2-$z$), and $spd$ for Si, C (diamond).
 ${\cal S}_{Q}$ stands for the charge spilling, and ${\cal S}_8$
for the spilling considering the first 8 bands.
The mean square error of the LCAO bands with respect to the PW
reference bands (in eV) is given for the valence bands ($\sigma_{v}$) and
two first conduction bands ($\sigma_{c}$).
Superscript $v$ means optimized for the valence.
STO 2-$z$ bases after Ref.~21.}
\vspace{2pt}
\begin{tabular}{lllllll}
&Basis & &${\cal S}_{Q}$ & ${\cal S}_{8(4)} $ & $\sigma_v$ & $\sigma_{c}$ \\
\hline \\
   & STO3$^{v}$&$sp$ & 0.0074 & 0.1548 & 0.29 & 1.95\\
   & STO& 2-$z$ & 0.0099 & 0.0735 & 0.36 & 1.50 \\
   & STO3$^{v}$&$spd$ & 0.0009 & 0.0101 &  0.09 & 0.23  \\
Si \\
   & PAO$_{5.0}$&$sp$ & 0.0139 & 0.1591 & 0.36 & 2.09   \\
   & PAO$_{5.0}$&2-$z$& 0.0085 & 0.0827 & 0.36 & 1.50  \\
   & PAO$_{5.0}$&$spd$ & 0.0042 & 0.0132 & 0.15 & 0.38 \\
\hline \\
   & STO3$^{v}$&$sp$ & 0.0024 & 0.0790 & 0.17 & 3.36 \\
   & STO&2-$z$ & 0.0032 & 0.0059 & 0.36 & 0.40 \\
C \\
   & PAO$_{4.0}$&$sp$ & 0.0041 & 0.0672 & 0.23 & 2.22   \\
   & PAO$_{4.0}$& 2-$z$ & 0.0018 & 0.0165 & 0.28 & 1.37 \\
\end {tabular}
\end{table}


\begin{table}
\caption{Band gap in Si and C (diamond) for single-$z$ and double-$z$
(2-$z$) bases. $E_{gap}^{\Gamma}$ and $E_{gap}$ stand for the direct
gap energy at $\Gamma$ and for the absolute gap energy, respectively,
both in eV. ${\bf k}$ is the wave vector of the minimum of the
conduction band in units of $\pi/a$, being $|{\bf k}|=1$ for the X point.}
\vspace{2pt}
\begin {tabular}{llllll}
&Basis & &  $E_{gap}^{\Gamma}$ & $E_{gap}$ & $|{\bf k}|$ \\
\hline
& STO3$^{v}$&$sp$ & 2.65 & 1.74 & 0.60 \\
& STO& 2-$z$ & 2.40 & 0.88 & 0.73 \\
& STO3$^{v}$&$spd$ & 2.58 & 0.76 & 0.85 \\
Si \\
 & PAO$_{5.0}$ & $sp$  & 2.47 & 2.18 & 0.47 \\
 & PAO$_{5.0}$ & 2-$z$  & 2.49 & 1.03 & 0.70 \\
 & PAO$_{5.0}$ & $spd$ & 2.65 & 0.91 & 0.85 \\ \\

 & PW  & & 2.55  & 0.45 & 0.85 \\
\hline
& STO3$^{v}$&$sp$ & 6.32 & 6.18 & 0.41 \\
& STO&2-$z$ & 5.46 & 4.30 & 0.73 \\
C \\
  & PAO$_{4.0}$ & $sp$ & 5.98 & 5.89 & 0.34 \\
  & PAO$_{4.0}$ & 2-$z$ & 6.00 & 5.28 & 0.53 \\ \\

  & PW & & 5.28 & 3.43 & 0.76 \\
\end {tabular}
\end{table}


\begin{table}
\caption{Calculated charge transfer using
optimized PAO$^{opt}_{\infty}$ basis
for some zincblende semiconductors.
$Q_C$ and $Q_A$ stand for the valence charge on the cation and the
anion, respectively, $\delta Q$ for the charge transfer with respect
to neutral atoms, and ${\cal S}_Q$ for the charge spilling. Numbers
in parenthesis were obtained from Hartree-Fock LCAO calculations,$^{22}$
except for GaAs, for which LDA LCAO was used.$^{23}$}
\vspace{2pt}
\begin{tabular}{llllll}
& Basis & ${\cal S}_Q$  & $Q_C$ & $Q_A$ & $\delta Q$ \\
\hline
BN   & $sp$   & 0.0022 & 2.19 (2.14) & 5.81 (5.86) &  0.81  (0.86) \\
BP   & $sp$   & 0.0038 & 3.51 (3.34) & 4.49 (4.66) & --.51 (--.34) \\
AlP  & $sp$   & 0.0035 & 2.15 (2.20) & 5.85 (5.80) &  0.85  (0.80) \\
SiC  & $sp$   & 0.0071 & 2.30 (2.19) & 5.70 (5.81) &  1.70  (1.81) \\
GaAs & $sp$   & 0.0041 & 2.58        & 5.42        &  0.42         \\
     & $spd$ & 0.0010 & 2.78 (2.88) & 5.22 (5.12) &  0.22  (0.12) \\
\end{tabular}
\end{table}

\begin{table}
\caption{ Calculated excess charge, in units of $e$,
 for the upper and lower Si atom
in the dimer of a Si(100)-2$\times$1
surface. A minimal $sp$ basis has been
used in all cases. ${\cal S}_Q$ stands for the
charge spilling. PAO$_\infty^{opt}$ and STO1 have been optimized for
bulk silicon. STO1$_{1.75}$ is a STO1 basis with $\beta=1.75$ a.u.$^{-1}$
(Ref.~5).}
\vspace{2pt}
\begin{tabular}{llll}
 Basis & ${\cal S}_Q$  &  $\delta Q_{up}$ &  $\delta Q_{down}$ \\
\hline
PAO$_\infty$ & 0.0080 & -0.10 & +0.12 \\
PAO$_\infty^{opt}$ & 0.0078 & -0.10 & +0.13 \\
PAO$_{5.0}$ & 0.0139 & -0.13 & +0.10 \\ \\
STO1$_{1.75}$ & 0.0318 & -0.15 & +0.10 \\
STO1 & 0.0076 & -0.13 & +0.12 \\
\end{tabular}
\end{table}

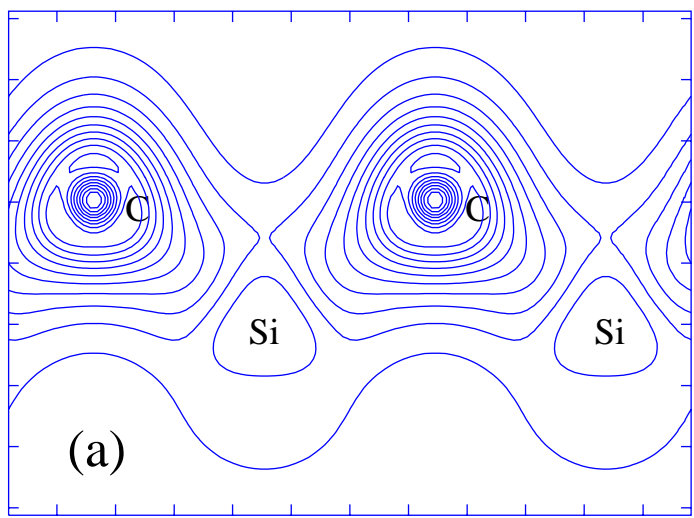
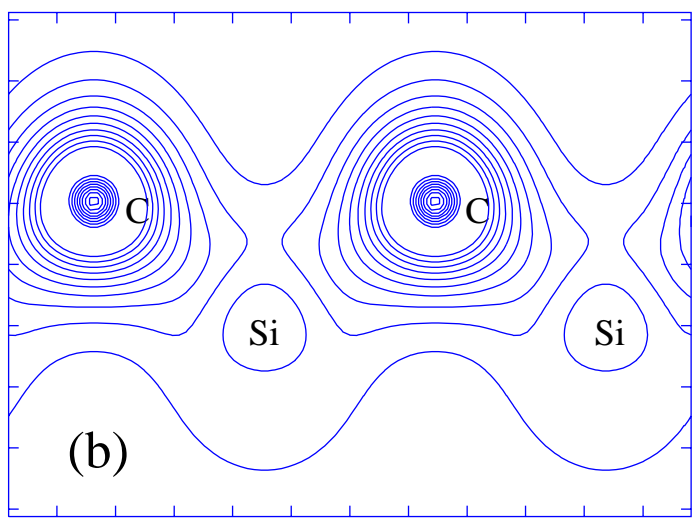
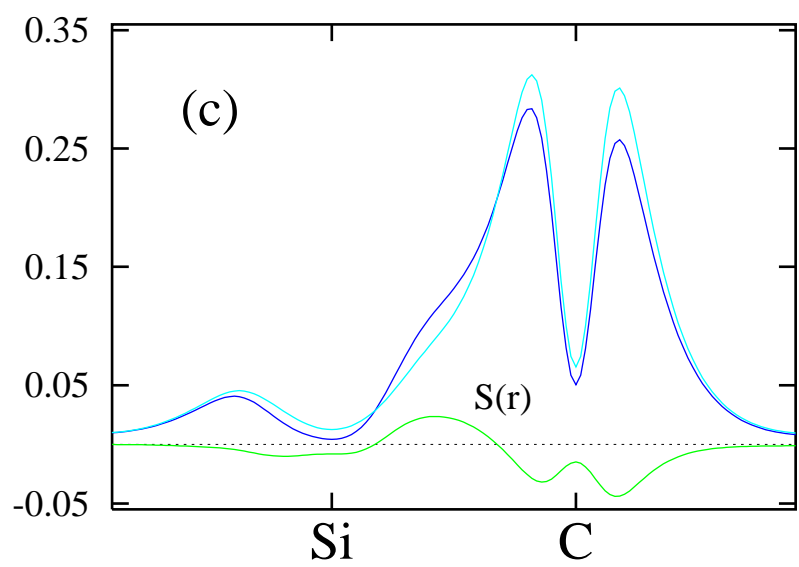
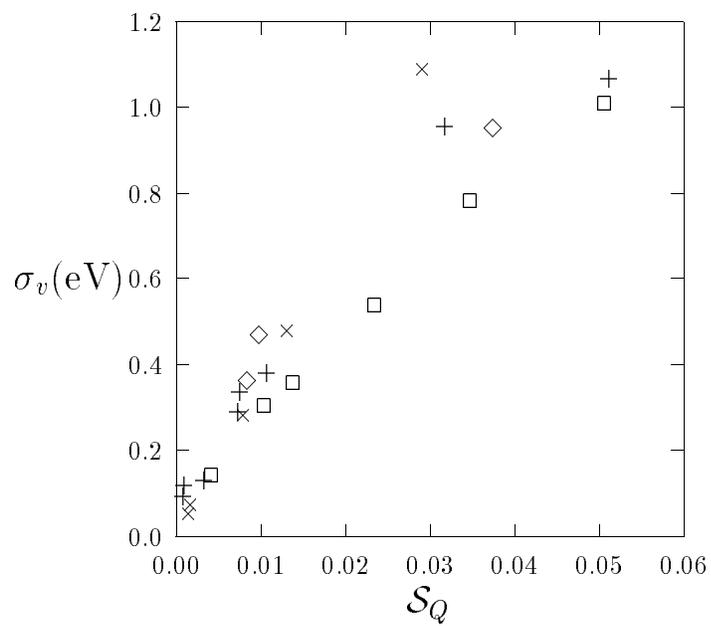
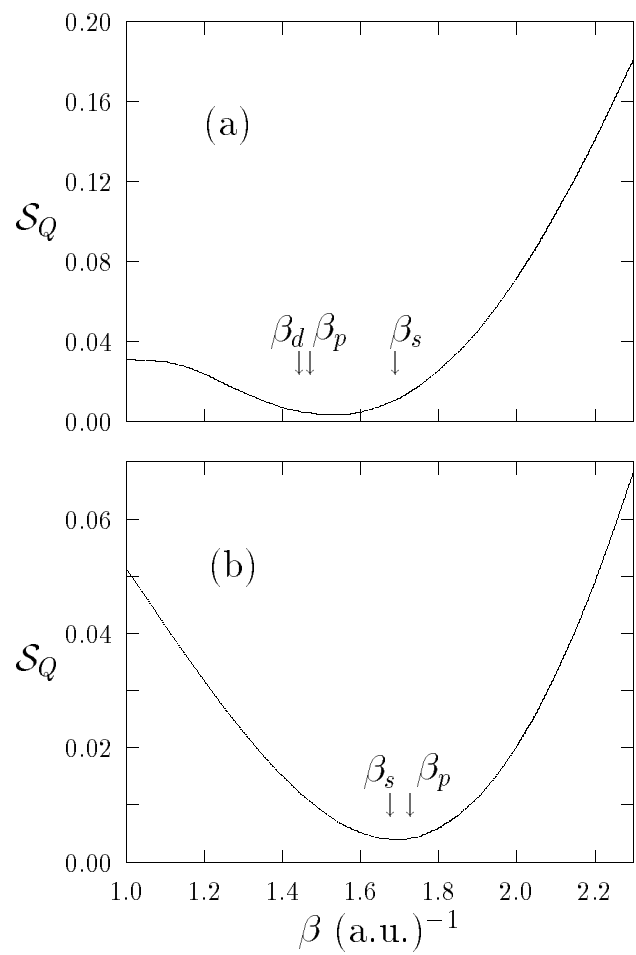
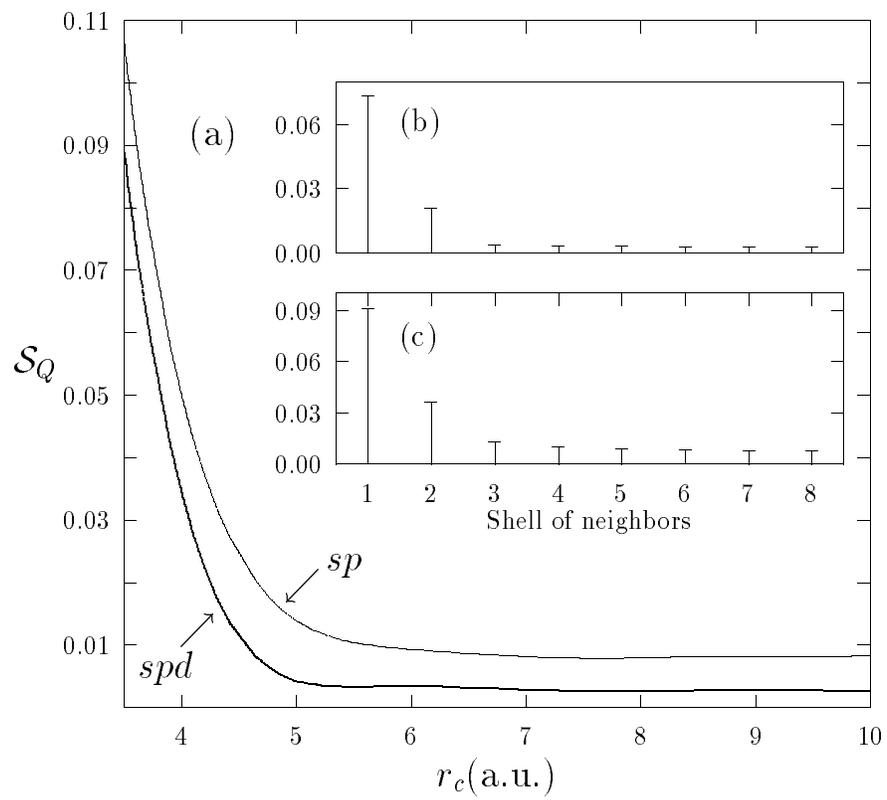
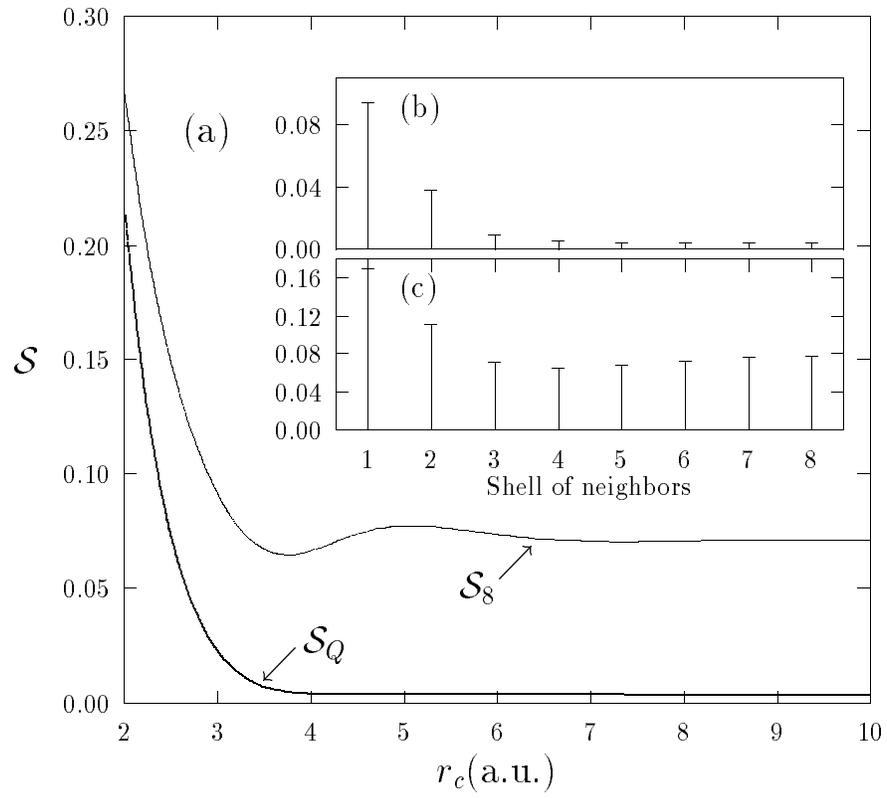
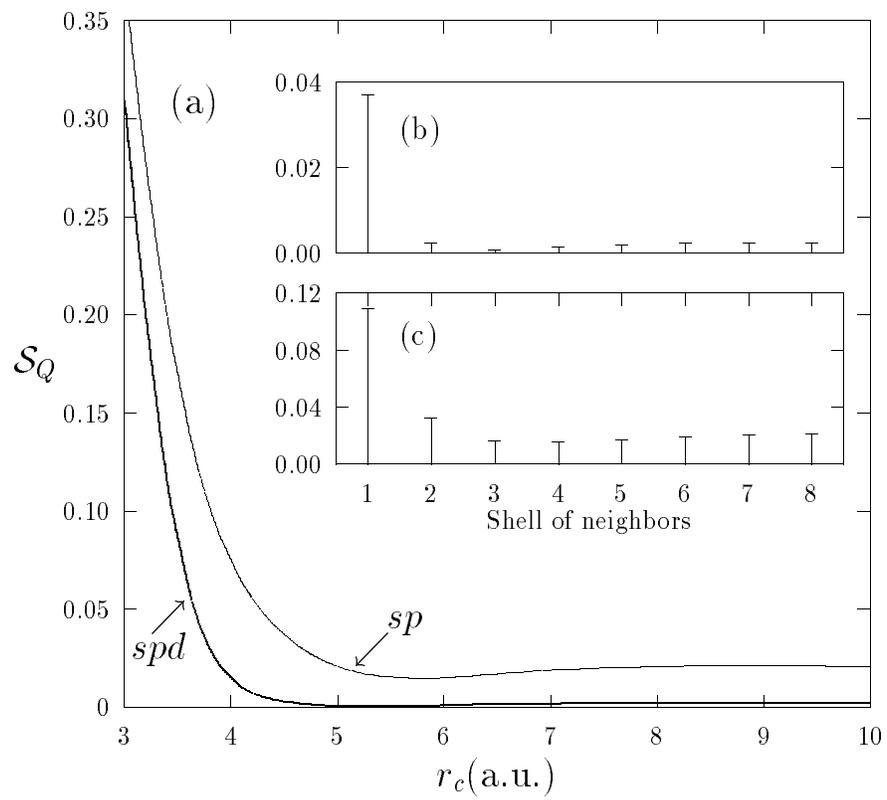
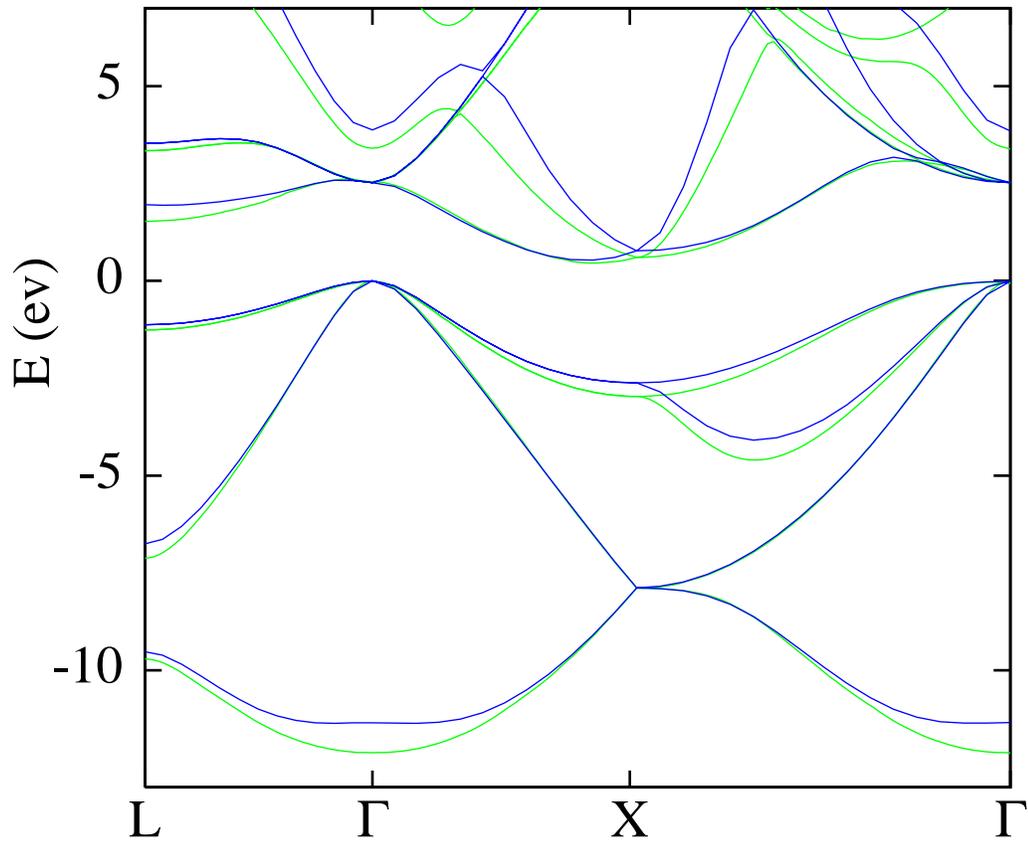
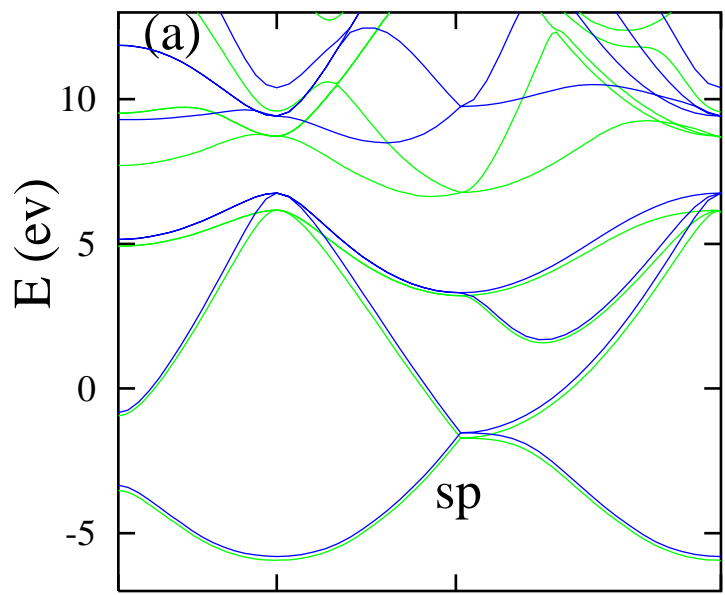
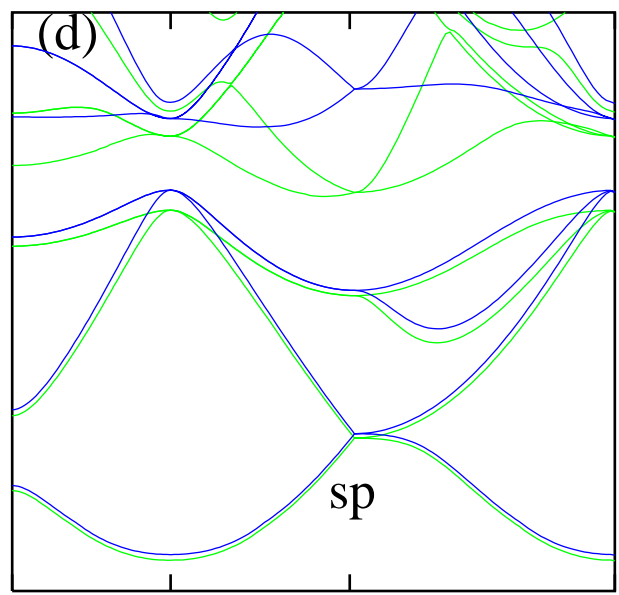
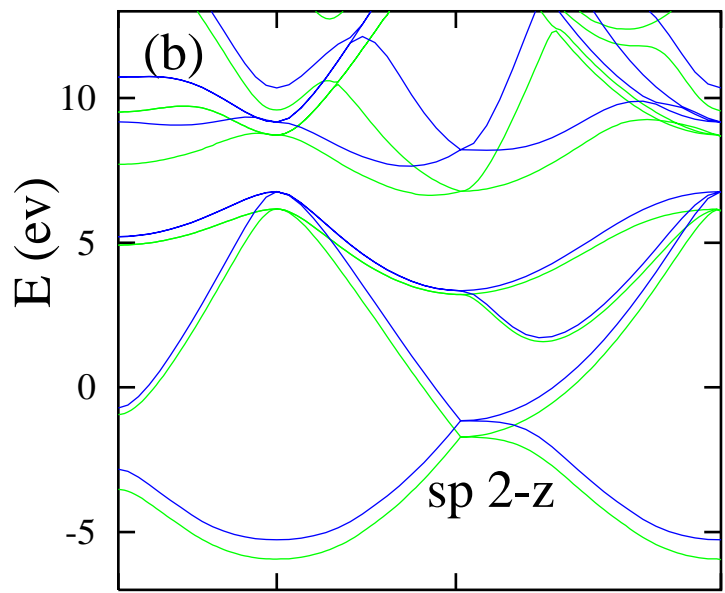
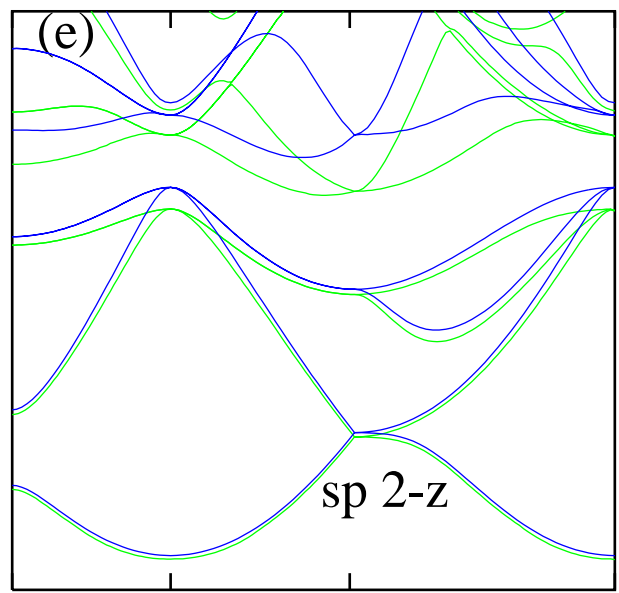
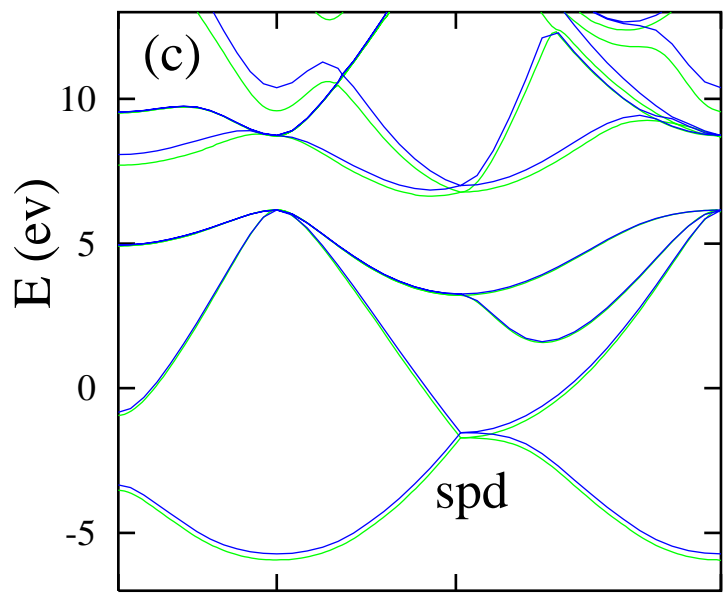
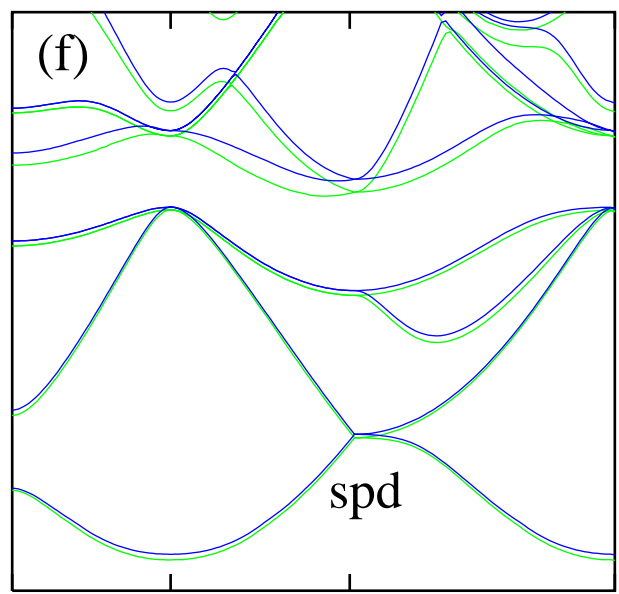
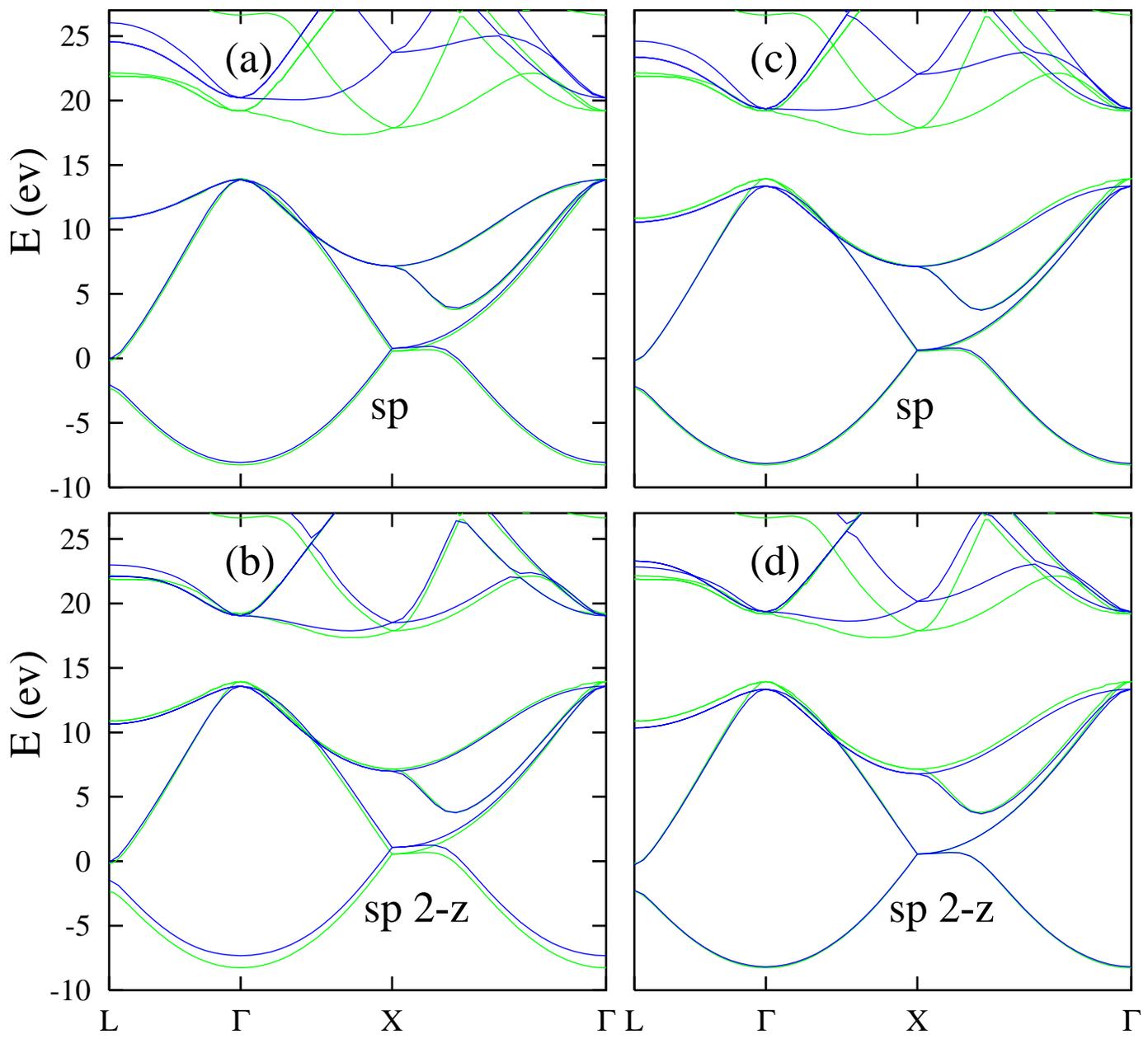
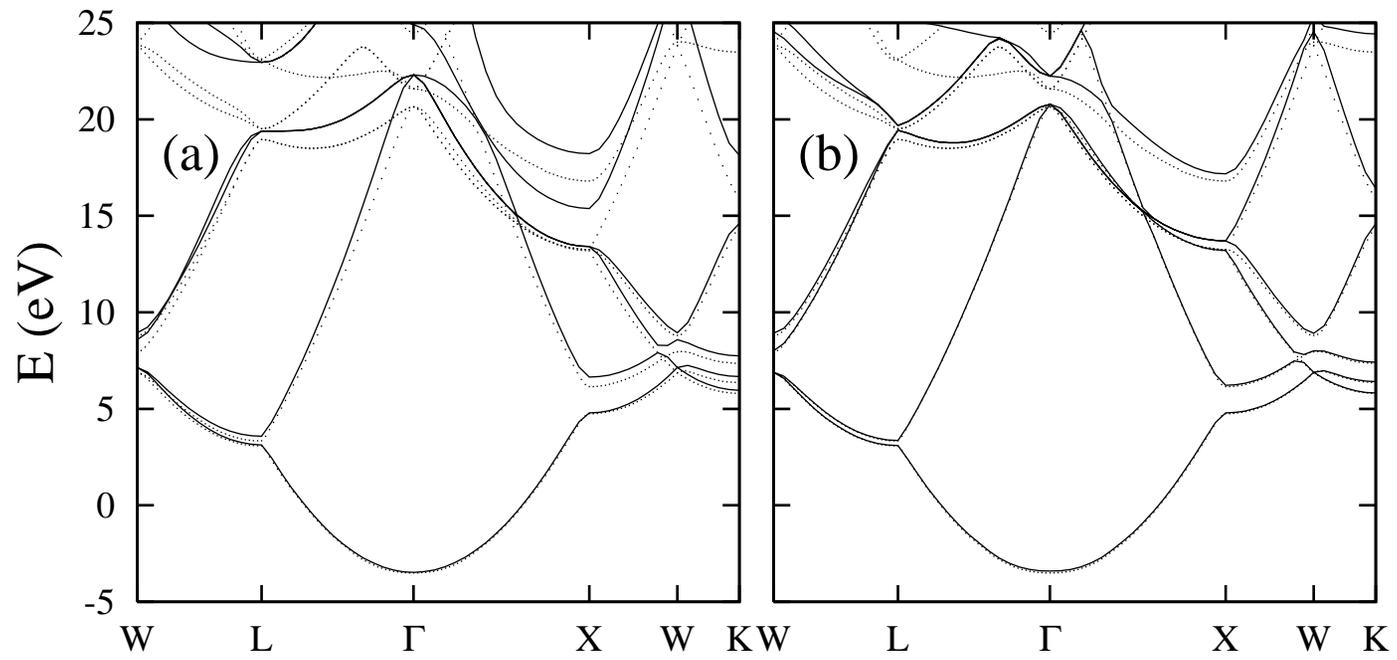
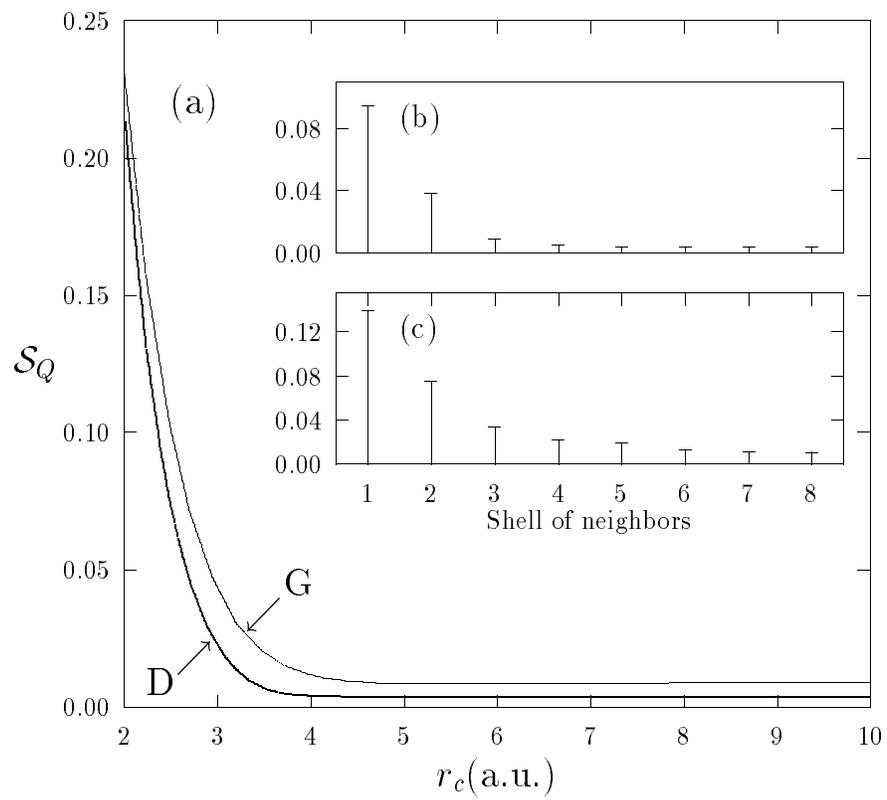
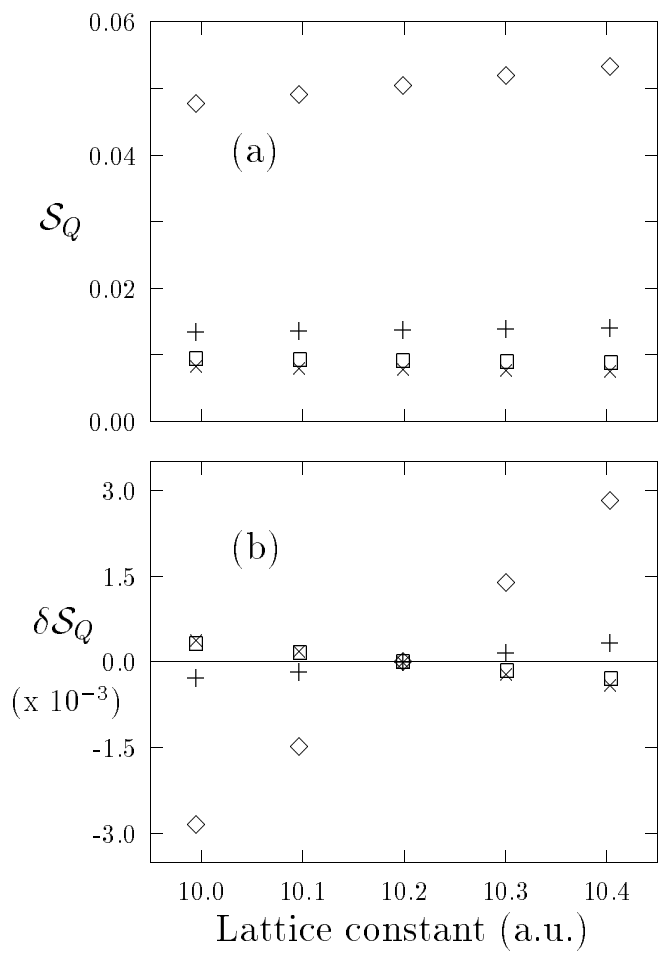
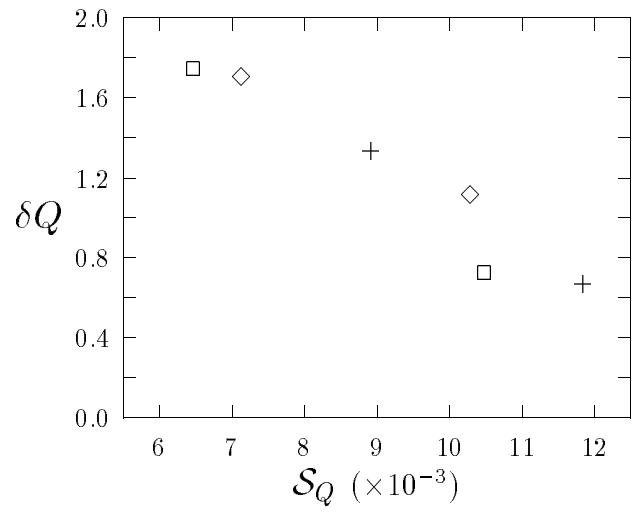
\end{document}